\documentclass[11pt]{article}
\pdfoutput=1

\usepackage{microtype}
\usepackage{authblk}
\usepackage{booktabs}
\usepackage{moreverb}
\usepackage{url}
\usepackage[numbers]{natbib}
\usepackage{listings}
\usepackage[colorlinks,bookmarksopen,bookmarksnumbered,citecolor=red,urlcolor=red]{hyperref}

\usepackage{pgfplotstable}
\pgfplotstableread{speeds.csv}{\SpeedTable}
\pgfplotstableread{bias.csv}{\BiasTable}

\begin{document}

\title{Fast keyed hash/pseudo-random function using SIMD multiply and permute}
\author{J.~Alakuijala}
\author{B.~Cox}
\author{\href{mailto:janwas@google.com}{J.~Wassenberg}}
\affil{Google Research}
\maketitle

\pdfinfo{
   /Author (J.~Alakuijala, B.~Cox, J.~Wassenberg)
   /Title (Fast keyed hash/pseudo-random function using SIMD multiply and permute)
   /CreationDate (D:20170206000000)
   /Subject ()
   /Keywords ()
   /Creator ()
   /Producer ()
}

\begin{abstract}

HighwayHash is a new pseudo-random function based on
SIMD multiply and permute instructions for thorough and fast hashing.
It is 5.2 times as fast as SipHash for 1 KiB inputs. An
open-source implementation is available under a permissive license.
We discuss design choices and provide statistical analysis,
speed measurements and preliminary cryptanalysis.
Assuming it withstands further analysis, strengthened variants may
also substantially accelerate file checksums and stream ciphers.

\end{abstract}


\section{Introduction}
\vspace{-2pt}

Hash functions are widely used for message authentication, efficient
searching and `random' decisions. When attackers are able to find collisions
(multiple inputs with the same hash result), they can mount denial of
service attacks or disturb applications expecting uniformly distributed
inputs. So-called `keyed-hash' functions prevent this by using a secret key to
ensure the outputs are unpredictable. These functions are constructed such that
an attacker who controls the inputs still cannot deduce the key, nor predict
future outputs. The authors of SipHash refer to these as
`strong pseudo-random functions' \cite{siphash}. However, existing approaches
are too slow for large-scale use. In this paper, we introduce two alternatives
that are about 2 and 5 times as fast as SipHash.

We are mainly interested in generating random numbers and message
authentication, for which 64-bit hashes are sufficient.
These are not `collision-resistant' because adversaries willing to
spend considerable CPU time could find a collision after
hashing about $\sqrt{\frac{\pi}{2}2^{64}}$ inputs.
However, small hashes decrease transmission overhead and are suitable for
authenticating short-lived messages such as network/RPC packets.
If needed, our approach can generate up to 256 hash bits at no extra cost.

Section~\ref{sec:related} briefly discusses existing hash functions and
their strength/speed tradeoff.
Section~\ref{sec:siptree} describes SipTreeHash, a j-lanes extension of
SipHash that is twice as fast on large inputs.
However, the SipHash construction is relatively slow because it
relies on rotate instructions not available in SIMD instruction sets.
Section~\ref{sec:highway} introduces HighwayHash, a novel hash function
that takes advantage of SIMD permute instructions for fast and thorough mixing.
Measurements in Section~\ref{sec:speed} indicate SipTreeHash is
twice as fast as SipHash for large inputs, and HighwayHash five times as fast.
Section~\ref{sec:security} describes our test suite and
shows that HighwayHash resists common attack techniques from the literature.
Further study may require new cryptanalysis techniques.
\section{Related Work} %
\label{sec:related} %
\vspace{-2pt}

Current cryptographic hashes require at least 2-3 CPU cycles per
byte~\cite{hashBenchmark}, which is about an order of magnitude slower than
fast hashes such as CityHash (0.23 c/b). Such a difference is unacceptable
to practitioners, especially if they are unconcerned about security.
SipHash~\cite{siphash} is a good compromise that has been studied
since 2012 without any known weaknesses. Our implementation requires about
1.25 c/b. Although relatively inexpensive for a strong hash, this is still
at least five times slower than fast hashes such as Murmur3 and CityHash.
However, these are vulnerable to collision and key extraction
attacks~\cite{cityWeak} and must not be exposed to untrusted inputs.
Several approaches have subsequently been proposed for taking advantage of
hardware-accelerated AES encryption~\cite{GCMSIV,aez}. These include security
proofs and are about twice as fast as SipHash.
The recent CLHash~\cite{clhash} is even faster despite requiring
1064 byte keys. However, its mixing is insufficient to pass smhasher's
avalanche test. A proposed fix adds an additional round of the ad-hoc
Murmur mixing function, but this still fails our distribution test for
zero-valued inputs (see Section~\ref{sec:smhasher}). Note that
CLHash was designed for speed and almost-universality, and is
not intended to withstand attacks \cite{clhashgoal}.

\noindent
We believe that SipHash remains a good default choice for non-cryptographic
applications because it offers (apparently) enough security at reasonable
speeds. A version with only 1 update and 3 finalization rounds is
1.2 to 1.9 times as fast (see Section~\ref{sec:speed}) while still
passing smhasher, which makes it an interesting candidate for
applications where security is less of a concern.

We develop two alternatives that further increase throughput
while retaining the simplicity and thorough mixing of SipHash.
We hope that these algorithms will replace unsafe hash functions and
increase the robustness of applications without incurring excessive CPU cost.
\section{SipTreeHash} %
\label{sec:siptree} %
\vspace{-2pt}

Maximizing performance on modern CPUs usually requires the use of
SIMD instructions. These apply the same operation (e.g.\ addition) to
multiple `lanes' (elements) of a vector. This works best for data-parallel
problems. However, the SipHash dependency chain offers limited parallelism and
cannot fully utilize the four AVX2 vector lanes.
Our SIMD implementation was actually slower than the scalar version,
presumably because of the lack of bit rotation instructions. We instead
compute four independent hash results by
logically partitioning input buffers into interleaved 64-bit pieces.
For example, consider a 64 byte input interpreted as eight 64-bit words:
$A_0, A_1, A_2, A_3, B_0, B_1, B_2, B_3$. These can be combined into four
hash results with two updates, one with $A_i$ and the other with $B_i$.
We must then fold the four results into a single hash. XOR reduction is
unsuitable because it cannot distinguish between permutations of the
64-bit words. Instead, we can just hash the results. This is known as a
tree-hash construction and has been used to accelerate SHA-256~\cite{jlanes}.
Being a straightforward extension of SipHash, this construction is likely
to be secure. However, its hash results are of course different, so this
cannot be used as a drop-in replacement.
Our implementation~\cite{highwaygithub} is available as open source
software under the Apache 2 license. We suggest HighwayHash be considered
instead because it is much faster, especially for smaller inputs.
\section{HighwayHash} %
\label{sec:highway} %
\vspace{-2pt}

Tree hash constructions can make efficient use of SIMD instructions.
However, the SipHash add-rotate-XOR construction is not ideally suited for
current instruction sets.
As previously mentioned, bit rotations must be implemented by ORing together
the result of left and right shifts. Although rotations of multiples of 8 bits
can be implemented with very fast byte permute instructions, this would weaken
SipHash to an unacceptable degree~\cite{siphash}.
The individual add and XOR instructions also only achieve a weak mixing effect.
By contrast, AVX2 includes $32 \times 32$ bit multiplication instructions that
mix their operands much more thoroughly. Although their latency is higher than
non-SIMD multiplies (5 vs 3 cycles~\cite{agner}), we believe the increase in
mixing efficiency vs.\ add/XOR is still worthwhile. The 64-bit Intel
architecture also provides 16 SIMD registers, which is enough to perform two
multiplies in parallel and thus hide some of the latency. Given that
multiplications are efficient, we now propose a new permutation step for
strong hashing.

\subsection{Zipper Merge}
Intuitively, it is clear that the highest and lowest bits of a multiplication
result are more predictable. This is the basis of the (problematic)
``middle square'' random generator~\cite{neumannRandom}, which only retains
the middle digits of a multiplication result.
We introduce a simple but seemingly novel approach: mixing multiplication
results with byte-level permute instructions.

Let us derive a suitable permutation. Recall that inputs are 64-bit
multiplication results that will become 32-bit multiplicands in the next
update round. It therefore makes sense to concentrate the poorly-distributed
top and bottom bytes in the upper 32 bits to ensure the multiplier bytes are
uniformly good (requirement 1).
To increase mixing, we also wish to interleave bytes from the neighboring
SIMD lane, ideally with no more than two adjacent bytes from the same lane
(requirement 2). Mostly importantly, we strive to equalize the `quality' of
each byte within a 64-bit lane (requirement 3).
We approximate this by counting the minimum distance of each bit's position
from the ends of its lane, computing the sum of these distances for the bits
in a byte, and sorting these sums in decreasing order. By this measure,
bytes 3 and 4 are best, 2, 5, 1, 6 are adequate, and 0 and 7 are worst.
Permutation results $R_i$ for a 16 byte lane pair $S_i$ are expressed as
hexadecimal offsets $P_i$, such that $R_i = S_{P_i}$. In other words, the
i-th offset indicates which source byte to copy into the i-th result byte.
Let $i=15$ be listed first and $i=0$ last. A partial permutation satisfying
requirements 1 and 2 takes the form
\texttt{7 8 6 9 ? ? ? ? 0 F 1 E ? ? ? ?}. To see this, note that the
source lanes (offset divided by 8) alternate, and that the byte offsets
(modulo 8) are \texttt{7 0 6 1}, which are the worst as mentioned.
It remains to distribute bytes 2-5 and A-D between both lower 32-bit
lane halves.  \texttt{5 2 C 3} and \texttt{D A 4 B} is a feasible solution,
with no more than two adjacent bytes and exactly equal quality in both lanes.
The final permutation is \texttt{7 8 6 9 D A 4 B 0 F 1 E 5 2 C 3}.
If we partition the inputs into two 64-bit parts and view these as
`highway lanes', the bytes (nearly) alternate between these two lanes.
We therefore borrow the term `zipper merge' from the road context.
Note that this particular permutation is not necessarily optimal --
it would be interesting, but computationally expensive, to optimize it
based on the resulting hash bit bias (discussed in Section~\ref{sec:security}).

\subsection{Update}

The proposed \verb|Update| folds a 32 byte vector of inputs (\verb|packet|)
into the 1024-bit internal state by multiplying and permuting.
The state variables \verb|v0|, \verb|v1|, \verb|mul0| and \verb|mul1| are
partitioned into independent 64-bit lanes.
\begin{lstlisting}[frame=lines,numbers=none,morekeywords={V4x64U}]
	v1 += packet;
	v1 += mul0;
	mul0 ^= V4x64U(_mm256_mul_epu32(v1, v0 >> 32));
	v0 += mul1;
	mul1 ^= V4x64U(_mm256_mul_epu32(v0, v1 >> 32));
	v0 += ZipperMerge(v1);
	v1 += ZipperMerge(v0);
\end{lstlisting}

\noindent
\verb|V4x64U| is a vector class with overloaded operators, which are
more convenient than using SIMD intrinsics directly.
However, some operations such as the crucial \verb|_mm256_mul_epu32|
multiplication must be expressed using intrinsics because there is
no C equivalent. This multiplies the lower 32 bits of its operands,
yielding a 64-bit result.
Note the symmetrical structures, which resemble the butterfly portion of
Fast Fourier Transforms (FT). This is reasonable -- recursive discrete FT
also need to combine or mix their inputs in a similar way.
The key operations here are permutations of the input and state,
which helps ensure no information about the key leaks (see the analysis in
Section~\ref{sec:security}).
Note that the multiplication latency is partially hidden behind other
instructions because \verb|mul| are only needed in the next \verb|Update|.
There is no need to similarly delay \verb|ZipperMerge| because its
latency is just one cycle~\cite{agner}.

\subsection{Finalization}
\label{sec:finalization}

After consuming all input data, the internal state must be further mixed to
reduce the risk of key leakage. It is convenient to use the same \verb|Update|
function. How many rounds are necessary? We hashed all $2^{24}$ combinations of
three input bytes and computed the probability of an output bit toggling
in response to an input bit changing. Strong biases remain after two rounds,
but decrease by a factor of 300 after three rounds to 0.03\%.
We added a fourth round for safety, which did not have a measurable impact
on the average bias.

To ensure the upper vector lanes are mixed into the final result, we must also
permute them between each round. Note that \verb|ZipperMerge| only
combines adjacent lanes. The AVX2 instruction set generally does not allow
interaction between the 128-bit halves of a vector, presumably to allow reuse of
logic from prior 128-bit SSE2 hardware. However, 64-bit lanes can be
shuffled at modest latency cost via \verb|_mm256_permutevar8x32_epi32|.
We swap the 128-bit vector halves and
also 32-bit lane halves of \verb|v0| using the following indices of
32-bit parts: \texttt{2 3 0 1 6 7 4 5}. The result is passed to \verb|Update|
and this process repeated another three times.

We then add together the four state vectors \verb|v*| and \verb|mul*|,
reducing the 1024-bit state into four 64-bit lanes. Mixing via addition
is slightly more thorough than XOR because of carry ripples.
A 64-bit hash suffices for many applications, in which case we only
retain the lower lane, which is slightly easier to extract into
general-purpose registers. Note that a quantum algorithm can
find collisions in $O(\sqrt[3]{N})$ time \cite{quantumCollision}.
If a hash is to resist such attacks, it should consist of at least
192 bits ($N=2^{192}$). HighwayHash can produce up to 256-bit hashes
at no additional cost; they also pass our test suite. 
However, our analysis of differential attacks in
Section~\ref{sec:differential} only applies to the 64-bit case.
HighwayHash needs further strengthening and analysis before it can
serve as a cryptographically secure message digest.

\subsection{Initialization}

We assume 256 key bits are reasonable and sufficient. Attackers have an
astronomically low, 1 in $2^{256-s}$ chance of guessing the key after
evaluating $2^s$ inputs.
The key must be expanded fourfold to populate the 1024 bits of internal state.
A two-fold expansion is achieved by initializing \verb|v0| with
the key XOR a constant, and \verb|v1| with the key (each lane rotated by
32 bits) XOR a second constant.
To make clear that there is no malicious intent behind the choice of
constants, we use ``nothing up my sleeve'' numbers and document the process.
We begin with a hexadecimal representation of the initial digits of
$\pi$~\cite{numberworld}.
To ensure each bit is set in at least one of the four lanes, we modify the
fourth number by setting each bit if zero or one other lanes have that bit set.
Finally, we initialize \verb|mul0, mul1| to the first and second constant,
respectively.

\subsection{Padding}
\label{sec:padding}

\verb|Update| operates on entire 32-byte vectors, so inputs must be
padded to multiples of that size. It is important that zero-valued
buffers of various sizes yield different hash results. The only
difference is their length, so a common approach is to inject an
encoding of the length into the input data. This can require another
mixing round if the input was a multiple of the vector size.
To minimize the likelihood of this happening, SipHash~\cite{siphash}
inserts only one byte encoding the size modulo 256. We instead suggest
updating the state directly, with a similar result as if the length had
been inserted into the input data. This avoids the overhead of
splicing a length code into a SIMD vector, which can be costly.
However, requiring the total size is burdensome for applications that
incrementally hash a stream of data. It should not be necessary to
maintain a separate counter in addition to the 1024-bit internal state.
We therefore inject only the size modulo 32 because size / 32 does
not add much information, especially compared to the mixing that
already occurs from every HighwayHash Update.
To accelerate avalanching of these five bits, we broadcast them into
each of the eight 32-bit vector lanes before adding to \verb|v0|.
For the same reason, we also permute \verb|v1| based on the size.
One convenient option is to rotate the 32-bit lanes by this amount.
Given the lack of rotate instructions, we resort to two shifts,
which is not a major cost because it only happens once before
finalization.

Padding would be much simpler if we could read past the end of the
input buffer, but this can trigger page faults or runtime safety checks.
Conceptually, we need to copy the remaining 0-31 bytes into a buffer,
load it into a vector and then \verb|Update|.
An architectural limitation motivates a slightly different approach. Stores to
memory or even cache involve considerable latency. To speed up subsequent
loads, CPUs forward data directly from intermediate store buffers.
However, smaller unaligned stores (from copying the remainder bytes) cannot be
combined to satisfy a larger load of the entire vector~\cite{intelOpt}.
We instead use the new AVX2 instruction \verb|_mm_maskload_epi32|
to load multiples of four bytes. The remaining zero to three bytes are
loaded individually, using only a single conditional branch, and inserted
into the packet. AVX2 requires the insertion position to be a compile-time
constant, so we insert into the most-significant four bytes.

\subsection{Source Code}

Our HighwayHash implementation \cite{highwaygithub} was published
as open source software in March 2016 under the Apache 2 license.
Its SIMD operations are expressed using \verb|V4x64U|, a custom AVX2
vector class with overloaded operators. For example,
\verb|V4x64U a = b + c| is easier to understand than the
equivalent SIMD intrinsic \verb|__m256i a = _mm256_add_epi64(b, c)|.
To support older CPUs, we also provide a SSE4.1 variant which is
about equally fast for small inputs, with about 80\%
of the AVX2 throughput for large inputs. An efficient dispatch
mechanism called \verb|InstructionSets| checks the current CPU
capabilities and jumps to the appropriate implementation. These are
packaged as template specializations, which avoids the overhead of a
virtual function call. This also enables HighwayHash to be embedded in a
larger inlined block of processor-specific code, with the dispatcher hoisted
out of any loops to minimize its overhead. Note that SIMD intrinsics
require an \verb|-mavx2| flag, which allows the compiler to generate AVX2
code everywhere. This can lead to crashes if inline functions are compiled
with different flags in multiple translation units.
Our \verb|InstructionSets| scheme avoids this by compiling SIMD code
in separate translation units, taking great care to avoid inline functions
in their shared headers. This implies that the dispatch overhead includes
a direct jump, which is inexpensive and only occurs infrequently if
callers are able to dispatch at a high level (outside of the inner loop).

\section{Throughput} %
\label{sec:speed} %
\vspace{-2pt}

We took unusual care in measuring the throughput of the
hash functions. SipHash, SipTreeHash and HighwayHash are implemented
in C++, the latter two using AVX2 intrinsics, and compiled by a
patched version of GCC 4.9 with flags {\small\texttt{-std=c++11 -O3
-march=haswell -fno-exceptions -fno-tree-vrp -fno-omit-frame-pointer 
-fno-strict-aliasing -fPIE}}.
Note that building with unmodified GCC 6.3 leads to about 10\% lower
throughput, whereas the 4.8.4 release results in three times slower
code. This is apparently caused by frequently spilling vector registers
to memory. It is unclear why the register allocations of the 6.3 and
patched 4.9 compilers are more efficient.

The benchmark runs on a single core of a desktop Xeon E5-2690 v3 clocked at
2.6~GHz. We record high-resolution timestamps from the invariant TSC and
use fences to ensure the measured code is not reordered by the
compiler or CPU.
To prevent elision of the benchmark computations, we pass the hash result as
input constraints to an empty inline assembly block. This forces the compiler
to assume that the result is used because the block allegedly modifies memory.
Prior measurements included the cost of initializing the 1~KiB input buffers.
This has been greatly reduced by only initializing the first 8 bytes.

We avoid unrealistic branch prediction rates by randomly interleaving
measurements for various sizes, rather than repeatedly hashing the same
input size.
Note that the benchmark frequently accesses the same data, which ensures
it is cache-resident, so the resulting throughput is an upper bound.
However, this is common in benchmarks and can be ignored because the
single-core hash throughput is lower than the observed memory bandwidth,
and far below the peak bandwidth~\cite{intelArk}.
To eliminate any outliers due to background activity or thermal throttling,
we use a robust estimator (the mode). The resulting measurements have a
mean absolute deviation of about 0.2 cycles; we retain the median as the
final result.
This `nanobenchmark' performance measuring infrastructure is included
in the open-source release \cite{highwaygithub}.

Table~\ref{tab:speed} lists throughputs for several input sizes in
the usual unit of CPU cycles per byte.
For 1~KB inputs, SipTreeHash and HighwayHash are 2.2
and 5.3 times as fast as SipHash. Reducing SipHash rounds from
2 per update and 4 during finalization to 1 and 3 also increases its
throughput by a factor of 1.2 to 1.9. HighwayHash is 1.4 to 1.9 times
as fast as SipTreeHash13.

\begin{table}[h]
\caption{Cycles per byte for various input sizes [bytes].}
\label{tab:speed}
\centering
\begin{tabular}{r|r|r|r|r|r|r}
\toprule
Algorithm & 8 & 31 & 32 & 63 & 64 & 1024\\
\midrule
SipHash          & 8.13  & 2.58 & 2.73 & 1.87 & 1.93 & 1.26\\
SipHash13        & 6.96  & 2.09 & 2.12 & 1.32 & 1.33 & 0.68\\
SipTreeHash      & 16.51 & 4.57 & 4.09 & 2.22 & 2.29 & 0.57\\
SipTreeHash13    & 12.33 & 3.47 & 3.06 & 1.68 & 1.63 & 0.33\\
HighwayHashSSE41 & 8.00  & 2.11 & 1.75 & 1.13 & 0.96 & 0.30\\
HighwayHashAVX2  & 7.34  & 1.81 & 1.71 & 1.04 & 0.95 & 0.24\\
\bottomrule
\end{tabular}
\end{table}

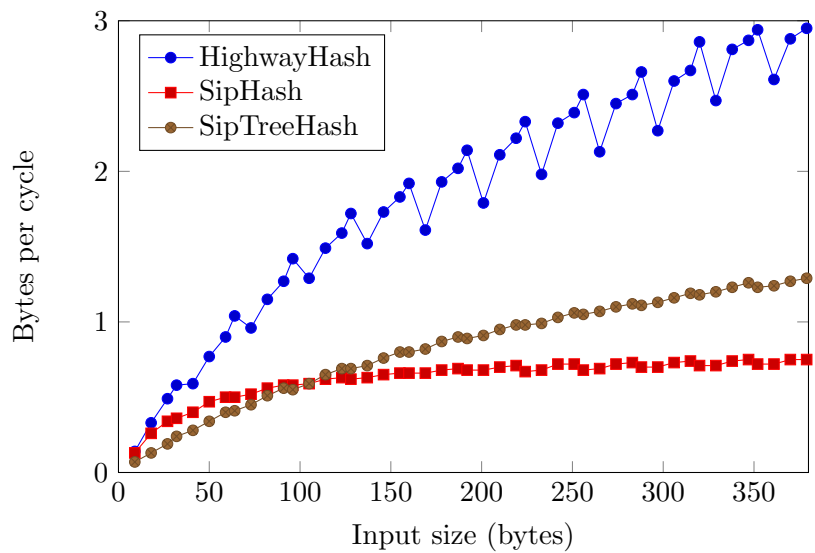
\begin{figure}
\centering
\begin{tikzpicture}
\begin{axis}[
  width=0.85\textwidth, height=0.6\textwidth,
  xmin=0, ymin=0, xmax=380,ymax=3,
  xlabel=Input size (bytes),
  ylabel=Bytes per cycle,
  legend style={cells={anchor=west}, legend pos=north west}
]
\addplot table [y=HighwayHash, x=Size] {\SpeedTable};
\addlegendentry{HighwayHash}
\addplot table [y=SipHash, x=Size] {\SpeedTable};
\addlegendentry{SipHash}
\addplot table [y=SipTreeHash, x=Size] {\SpeedTable};
\addlegendentry{SipTreeHash}
\end{axis}
\end{tikzpicture}
\caption{Throughput [bytes per cycle] increases with input size, but with
dips or peaks at multiples of the packet size. SipTreeHash is slower than
SipHash for small inputs but twice as fast for large inputs. HighwayHash is
faster for all sizes due to its optimized padding scheme.}
\label{fig:speed}
\end{figure}

\noindent
Throughput increases for larger inputs (Figure~\ref{fig:speed})
because the finalization cost is amortized over more data.
SipTreeHash is slower than SipHash for smaller inputs because it
processes entire 32-byte AVX2 vectors, and also hashes the
intermediate results.
However, AVX2 SIMD instructions can process four 64-bit elements at a
time, so SipTreeHash eventually overtakes SipHash. It would be
faster still if SIMD instruction sets supported bit rotations, which
are currently emulated with three instructions.
Note the periodic decreases at multiples of 32 bytes in both tree hashes.
Inputs are padded to entire vectors, so relative throughput
decreases as padding increases. To capture this, we measured the
throughput at sizes $32 \cdot i + \{0, 9, 18, 27\}$, which covers
all four possible values of size modulo 4. For HighwayHash, the
best case is a multiple of 32, which requires no padding.
SipHash was designed to efficiently handle small inputs and is
less affected by input size and padding.
The tree hashes are much faster for large inputs because they
process 32 bytes at a time. Surprisingly, HighwayHash outperforms
SipHash for small inputs due to the optimized padding described
in Section~\ref{sec:padding}.

\section{Security Analysis}%
\label{sec:security}%
\vspace{-2pt}

We are not experienced cryptographers, and do not see a way to reduce the
HighwayHash algorithm to provably secure constructions.
This multiply-permute scheme may require new methods of cryptanalysis.
In the absence of a formal proof, we rely on statistical testing to
validate our main claim: HighwayHash is a keyed hash that is
indistinguishable from a uniformly random source, with an avalanche
effect comparable to a cryptographically random number generator.

\begin{figure}
\centering
\begin{tikzpicture}
\begin{axis}[
  width=0.85\textwidth, height=0.6\textwidth,
  xmin=4, ymin=0.064, xmax=32,ymax=0.076,
  xlabel=Input size (bytes),
  ylabel=Median bias (\%),
  legend style={cells={anchor=east}, legend pos=south east}
]
\addplot table [y=HighwayHash, x=Size] {\BiasTable};
\addlegendentry{HighwayHash}
\addplot table [y=SipHash, x=Size] {\BiasTable};
\addlegendentry{SipHash}
\addplot table [y=RDRAND, x=Size] {\BiasTable};
\addlegendentry{RDRAND}
\end{axis}
\end{tikzpicture}
\caption{Median avalanche bias from 49 independent trials with
random inputs and keys. Both SipHash and HighwayHash appear
to have avalanching properties indistinguishable from the
RDRAND cryptographically secure random generator according to
this test.}
\label{fig:bias}%
\end{figure}
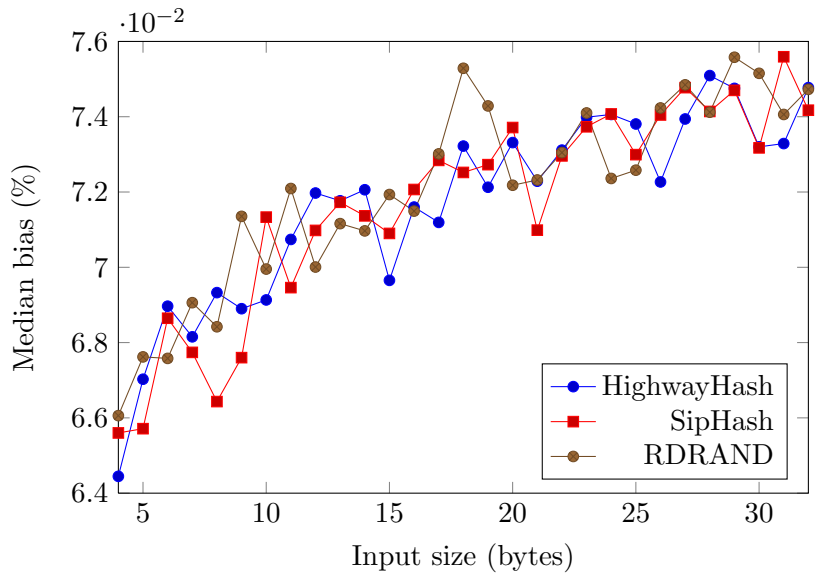

\subsection{smhasher Test Suite}
\label{sec:smhasher}%

smhasher \cite{smhasher} is a test suite for hash functions that
verifies their output distribution and checks for collisions when
hashing `difficult' inputs.
Recall that CLHash (without additional mixing) fails the
avalanche test \cite{clhash}, which requires that half of the
output bits change when an input bit is flipped. We further
strengthen this test by checking all input sizes between 4 and
32 bytes\footnote{Biases computed for 3 bytes are much larger and
likely overestimates.} and raising the iteration count from
300K to 30M. Previously, smhasher reported the resulting
maximum bias for each size. However, these are prone to outliers.
To prevent this, we draw 49 bias samples using different input and
hash seeds, and retain the median. We also improve the quality of
the random inputs. A generator with period $p$ should not be asked to
produce more than $\sqrt[3]{p}$ numbers \cite{randomBirthday}.
The input data for all sizes are populated from 78 64-bit random numbers.
This implies a minimum period of $2^{116}$, which rules out
common generators such as XorShift64 and Tausworthe ($2^{88}$).
We instead use the \verb|pcg64_k32| permuted congruential
generator \cite{randomPCG} with a much larger period of $2^{2176}$.
The avalanche test is considered successful if each output bit has a
bias (deviation from the expected 50\% bit flip rate) of less than 1\%.
FarmHash 1.1's \texttt{Fingerprint64}\footnote{\texttt{Hash64WithSeed}
mixes more thoroughly, but its API contract does not guarantee
unchanging results.} \cite{farmhash}, a recent fast non-cryptographic
hash, results in 62-77\% bias for 25-32 byte inputs.
This may enable key recovery or collision attacks. The other hashes
have less than 0.08\% bias (Figure~\ref{fig:bias}). To estimate a
lower bound, we also computed the bias of a high-quality random
number generator masquerading as a hash function. Intel's
RDRAND is a cryptographically secure generator based on an
AES block cipher, periodically reseeded from observations of
thermal noise. Interestingly, the SipHash and HighwayHash biases
are indistinguishable from that of the hardware random generator.
We can only conclude that all tested hash functions appear
to mix equally thoroughly according to this test.%
In addition to empirical verification, we discuss several possible attacks and
whether they apply to HighwayHash. Our attack model assumes that
the secret key is initially unknown.

\subsection{Differential Attack}
\label{sec:differential}
Let us begin by attempting a differential attack. Assume we start in a
randomized state, i.e.\ \verb|v0, v1| are uniformly distributed random
variables, which is holds true if the key was random. We will attempt to
modify sequential vectors A, B and C such that the state change caused by A is
reversed by B.  The two rounds of \texttt{Update} have the following effect:
\begin{lstlisting}[frame=lines,morekeywords={V4x64U}]
	v1 += A;
	v1 += mul0;
	mul0 ^= V4x64U(_mm256_mul_epu32(v1, v0 >> 32));
	v0 += mul1;
	mul1 ^= V4x64U(_mm256_mul_epu32(v0, v1 >> 32));
	v0 += ZipperMerge(v1);
	v1 += ZipperMerge(v0);
	v1 += B;
	v1 += mul0;
	mul0 ^= V4x64U(_mm256_mul_epu32(v1, v0 >> 32));
	v0 += mul1;
	mul1 ^= V4x64U(_mm256_mul_epu32(v0, v1 >> 32));
	v0 += ZipperMerge(v1);
	v1 += ZipperMerge(v0);
	v1 += C;
\end{lstlisting}
The last step is included in case C can help restore the value of \texttt{v1}.
The effect of A is to either change \texttt{mul0} in line 3, or \texttt{mul1} in
line 5, except for a 1 in $2^{32}$ chance that the 32 bit part of \texttt{v0}
is 0.  In that case, \texttt{v1} can probably be corrected in line 8, but we
have still changed \texttt{v0}, which cannot be corrected before propagating
to \texttt{mul0} or \texttt{mul1} with all but 1 in $2^{32}$ chance, and the
differential attack fails because we are no longer able to influence
\texttt{mul0} or \texttt{mul1}.  The combined probability that both changes
from A and B do not affect either \texttt{mul} value is at most 1 in $2^{64}$,
so that approach also fails.
The other case is where we allow \texttt{mul0} or \texttt{mul1} to change in
line 3 or 5, hoping it will be corrected by the corresponding line 10 or 12.
The probability of this is $2^{-32}$, with a higher likelihood when we
change only 1 bit in A and then B, and lower when we change more bits.
In line 3 or 5, there will be at least a 32-bit random change in either
\texttt{mul0} or \texttt{mul1}, which propagates to \texttt{v0} or
\texttt{v1} in line 9 or 11.  Correcting \texttt{v0} or \texttt{v1} is
at best another 1 in $2^{32}$ chance in lines 13-15, and again the
attack fails. Note that this analysis holds even in the first vector where the
\texttt{mul0} and \texttt{mul1} values are known, but \texttt{v0} and
\texttt{v1} are random. Longer differentials that include a subsequent vector
are too late to help because changing one or more bit in A will avalanche from
adding/XORing/multiplying at least 64 bits of random initial state at line 1
from \texttt{v0 v1 mul0 mul1} into each other after line 11.

\subsection{Length Extension Attacks}
Length extension attacks are infeasible because the calls to
\texttt{PermuteAndUpdate} during finalization differ from previous
calls to \texttt{Update}.  In particular, the permutation involves neighboring
lanes, whereas regular \texttt{Update} invocations do not. Also, the state is
random, depending on the secret key, and permutes in a manner that avalanches to
new states on each call to \texttt{Update}, regardless of the vector values.
In cases where the input length is not a multiple of the vector size (32 bytes),
the padding scheme ensures that messages with various numbers of trailing
zero bytes result in different hashes.

\subsection{Entropy Loss}
When all input vectors except one are held constant, the final state will differ
for all $2^{256}$ possible vector values, because \texttt{Update} is a
permutation. Compression only occurs when injecting input data and
discarding all but the 64 output bits during finalization.

\subsection{Rotational Attacks}

The security of an ARX (Add, Rotate, XOR) scheme $S$ depends on the number of
additions $q$ \cite{arxRotate}.
For $n$-bit inputs $I$ and a rotation function $R$ we have
\[Z = P(R(x + y, r) = R(x, r) + R(y, r)) = (1 + 2^{r-n} + 2^{-r} + 2^{-n}).\]
\[P[S(R(I, r)) = R(S(I), r)] = Z^q\]

\noindent
For an ARX scheme with too few additions, we can detect non-randomness in the
function by showing that $S(R(I, r)) = R(S(I), r)$ more than a random function,
the probability of which is $2^{-n}$ for an $n$-bit hash function. For example,
SipHash (with 2 update and 4 finalization rounds) involves 13 additions.
The resulting state values, when XORed together, will be the same whether we
rotate the inputs by 32 or just the output by 32 with probability
$4^{-13} = 2^{-26}$. If keys are reused and attackers choose the input,
a significant bias should be detectable. However, it is not clear how this can
be used to either create collisions or reveal the key. Direct rotational
attacks to not seem to apply to the multiplier, since the output is 64-bits
whereas each operand is 32 bits.  However, each multiplier contributes on the
average 16 32-bit additions, and the four finalization rounds include
8 multiplications for a total average of 128 32-bit additions. For the results
to be equal after rotating both inputs by 16, we need the upper 16 bits of one
operand to be 0, and the lower 16 bits of the other to be 0, with probability
$2^{-32}$.  The chance of this happening throughout all finalization rounds is
negligible.  With a total of 32 multiplications (4 lanes times 8), the
resulting 1 in $4^{1024}$ bias is far too small to detect.

\subsection{SAT Solvers}
Boolean satisfiability (SAT) solvers have been applied to the problem of
deriving unknown key bits when the attacker knows most bits of the key in
SipHash. Modified versions that use fewer finalization rounds were found
to be vulnerable to modern SAT solvers. However, formal verification of CPU
multiplier circuits remains a challenge. Polynomial time solutions for
verification have been discovered, but they rely on algebraic properties of
multiplication and cannot be used on larger circuits also including XOR gates.
We therefore expect HighwayHash to be far more resistant to SAT-solver
based key extraction than traditional ARX hash functions such as SipHash.

\section{Conclusion}
\vspace{-2pt}

Faster hashing could save enormous amounts of CPU time in data centers.
However, algorithms entirely focused on speed, including Murmur3 and
CityHash, are vulnerable to attacks \cite{cityWeak}. For example,
adversaries can skew the distribution of `random' decisions. We describe a
strengthened version of the smhasher test suite \cite{smhasher} that reveals
a previously unknown weakness in \texttt{Fingerprint64} from
FarmHash 1.1 \cite{farmhash} (but not \texttt{Hash64WithSeed},
which is fast and well-distributed). 
Applications should not use such unsafe hash functions unless they
trust their input data. SipHash is a widely used pseudo-random
function for which no relevant attacks are known.
However, it is relatively slow (1.25~cycles per byte for 1~KB inputs).
We propose HighwayHash, a novel keyed hash that is about 5 times as fast
thanks to SIMD multiply and permute instructions.
The source code is available under the permissive Apache 2 license at
\url{https://github.com/google/highwayhash}.
To the best of our (non-expert) ability, we have analyzed the
algorithm and found no weaknesses. Statistical tests indicate the
HighwayHash output bits are as uniformly distributed as the
cryptographically secure RDRAND generator.
We also considered powerful attacks, including differential and rotational,
and are confident the algorithm will withstand these specific attacks.
We welcome further cryptanalysis -- the family of hash functions built
from multiplications and byte permutations had not been studied yet.
Assuming this construction remains unbroken, it appears useful for
pseudo-random number generators and fast message authentication codes.
Strengthened variants may also substantially accelerate stream ciphers and
file hashing.

{
\small
\bibliographystyle{unsrtnat}
\bibliography{references}{}
}

\end{document}